\documentclass[prl,noshowpacs,english,twocolumn]{revtex4}
\pdfoutput=1
\usepackage{amsmath,amssymb,amsfonts,mathrsfs}
\usepackage{amsthm}
\usepackage{CJK}
\usepackage{bm}
\usepackage{babel}
\usepackage{graphicx}
\allowdisplaybreaks
\usepackage{braket}

\begin{document}

\title{A potential superhard carbon allotrope: T$_{5}$-carbon}

\author{Xian-Yong Ding$^{1}$, Chao Zhang$^{1,3,5,\ast}$, Dong-Qi Wang$^{2}$, Bing-Sheng Li$^{3}$, Qingping Wang$^{1}$, Zhi Gen Yu$^{4}$, Kah-Wee Ang$^{5}$, and Yong-Wei Zhang$^{4}$}

\address{$^1$School of Materials Science and Engineering, Anhui University of Science and Technology, Huainan 232001, China}
\address{$^2$Multidisciplinary Initiative Center, Institute of High Energy Physics, Chinese Academy of Sciences, Beijing 100049, China}
\address{$^3$State Key Laboratory for Environment-friendly Energy Materials, Southwest University of Science and Technology, Mianyang, Sichuan 621010, China}
\address{$^4$Institute of High Performance Computing, A*STAR, Singapore, 138632}
\address{$^5$Department of Electrical and Computer Engineering, National University of Singapore, Singapore, 117583}

\begin{abstract}
A novel stable carbon allotrope is predicted by using first-principles calculations. This allotrope is obtained by replacing one of the two atoms in the primitive cell of diamond with a carbon tetrahedron, thus it contains five atoms in one primitive cell, termed T$_{5}$-carbon. The stabilities of T$_{5}$-carbon are checked in structural, thermal, vibrational and energy calculations. The calculations on electronic, thermal, and mechanical properties reveal that T$_{5}$-carbon is a semiconductor with an indirect band gap of 3.18 eV, and has a lattice thermal conductivity of 409 W/mK. More importantly, the Vickers hardness of T$_{5}$-carbon is 76.5 GPa which is lower than that of diamond but larger than those of T-carbon and cubic boron nitride, confirming the superhard properties of T$_{5}$-carbon, suggesting its wide applications in mechanical devices.
\end{abstract}


%Use showkeys class option if keyword %display desired

\maketitle

Carbon, one of the basic components of our universe, derives a variety of allotropes with a wide range of fascinating properties. This is mainly promoted by the diversiform hybridization between the C-$s$ and C-$p$ orbitals of carbon atoms. Over the past decades, with the improvement of computational science and capacity, a variety of carbon allotropes have been prdicted, such as, Penta-graphene\cite{PentaGraphene}, T-carbon\cite{T-carbon}, bco-C$_{16}$\cite{bcoc16}, D-carbon\cite{D-carbon} and so forth\cite{cR6,Rh6}. Meanwhile, as the synthesis technology is growing and maturing, a series of carbon phases are successfully synthesized, i.e., fullerences\cite{fullerene}, carbon nanotubes\cite{cnt}, graphene\cite{graphene} and T-carbon\cite{T-carbon-nonowire}. These synthesised carbon allotropes attach enormous impacts in the field of chemistry, physics, and materials, and greatly promote the advancement in nanotechnology and nanodevices. It seems that the applications of carbon or carbon-based materials would be the mainstream in the future.

Due to the character of $sp^{3}$ hybridization of carbon atoms, its allotropes have always been considered to be the superhard materials which exhibit superior mechanical performances. Thus, it has attracted special attention to researchers focusing on synthesising and predicting new carbon allotropes. Several novel superhard carbon allotropes, such as, the monoclinic M-carbon\cite{M-carbon}, bct-C$_{4}$\cite{bctc4}, W-carbon\cite{W-carbon} and O-carbon\cite{O-carbon} have been predicted to simulate the high-pressure phase\cite{SuperhardCarbon-exp}, whose hardness is even higher than that of diamond. Recently, some new predicted carbon allotropes Cco-C$_{8}$\cite{CcoC8} and C$_{20}$-T\cite{C20-T} have also been proposed with superhard properties. These predicted superhard carbon allotropes promote the advancement in understanding the mechanism of superhard materials.

In this work, using first-principles calculations, we obtain a novel carbon allotrope which possesses a higher Vicker hardness.
This allotrope containing five atoms in one primitive cell, can be obtained by substituting one of the two atoms in the primitive cell of diamond with a carbon tetrahedron, thus termed T$_{5}$-carbon. T$_{5}$-carbon exhibits considerable stabilities in mechanics, dynamics and thermodynamics. Furthermore, the predicted XRD spectra show that it has a contribution in the production of TNT and diesel oil detonation\cite{XRD-Soot2003}. Importantly, the calculated Vicker hardness can reach up to 76.5 GPa with a comparable smaller elastic modulus, which indicates that the wide application in mechanical devices.

First-principles calculations are implemented in Vienna $ab$ $initio$ simulation package \cite{Kresse1,Kresse2}. The core-valence electron interactions are solved using the projector augmented-wave method\cite{Blochl}. The cutoff energy of 600 eV for plane waves is used in all calculations. The generalized gradient approximation is introduced to describe the exchange-correlation functionals within the Perdew-Burke-Ernzerhof method \cite{Perdew}. The $\Gamma$-centered K-mesh of $16\times 16 \times 16$ is used for the ionic relaxation and electronic self-consistent iteration. The conjugate gradient method is used to relax the  atomic positions. The forces of all carbon atoms are converged to be less than $1.0\times10^{-3}$ eV/{\AA}. The harmonic (second order) interatomic force constants (IFCs) are calculated by employing the density functional perturbation (DFPT) method \cite{DFPT}. Then we obtain the phonon dispersion by using the PHONOPY code \cite{PHONOPY}. The Crystal Orbital Hamilton Populations (COHPs)\cite{COHP-1,COHP-2,COHP-3} of diamond, T$_{5}$-carbon and T-carbon are also implemented in the LOBSTER code\cite{lobster}. Furthermore, In order to calculate the lattice thermal conductivity $\kappa$, we acquire the anharmonic (third order) IFCs with considering the interaction of sixth nearest carbon neighbors. A supercell of 3 $\times$ 3 $\times$ 3 is used for the second order IFCs and third order IFCs calculations. Finally, the phonon Boltzmann transport equation (BTE) with iterative method are solved as implemented in the ShengBTE code\cite{ShengBTE2014}.
\begin{figure}\label{fig1}
\includegraphics[scale=0.46]{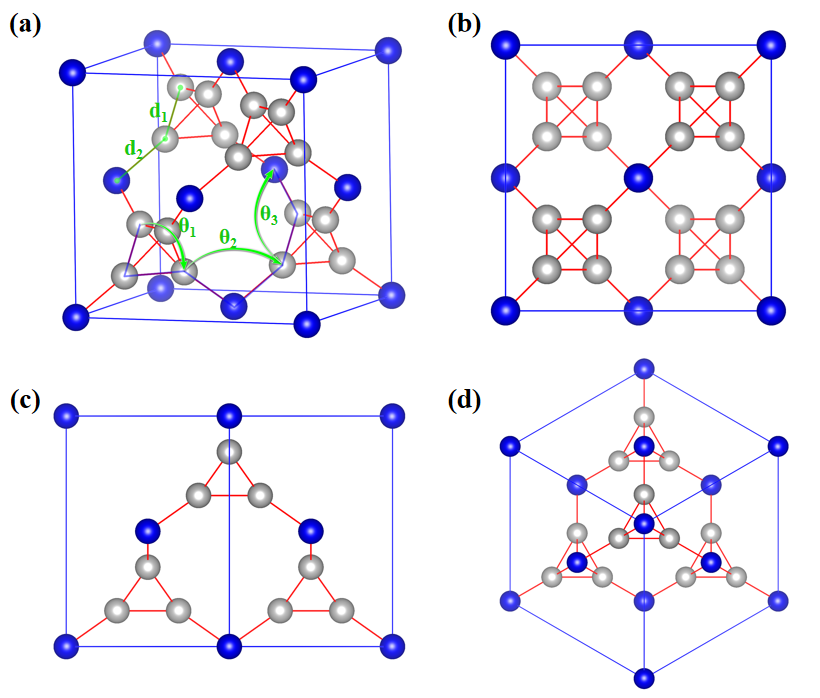}
\caption{(Color online) Crystal structure of T$_{5}$-carbon. (a) Cubic crystalline of T$_{5}$-carbon with the space group F$\overline{4}$3M (No. 216). Two different bond lengths ($d_{1}$, $d_{2}$) and three different bond angles ($\theta_{1}$, $\theta_{2}$, $\theta_{3}$)) are also listed. (b)-(d) Views from [100], [110], and [111] directions of T$_{5}$-carbon, respectively. Note that the blue and gray balls denote the two non-equivalent carbon atoms which occupy Wyckoff positions of 4a (0, 0, 0) and 16e (0.844, 0.844, 0.844), respectively.}
\end{figure}

\begin{figure}\label{fig2}
\centering
\includegraphics[scale=0.56]{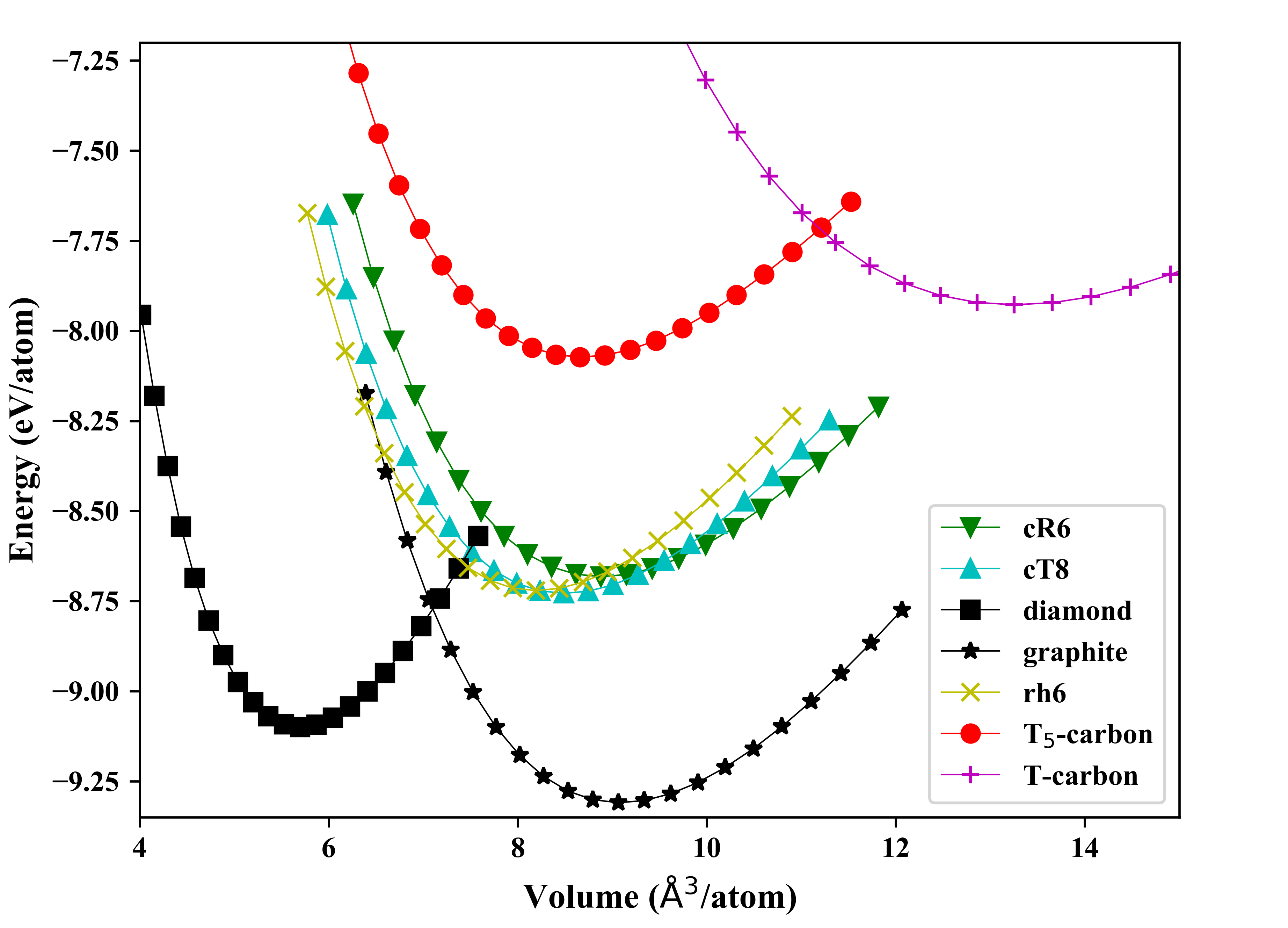}
\caption{(color online). Total energy per atom versus volume of T$_{5}$-carbon. Carbon allotropes of cR6, cT8, diamond, rh6, T-carbon and graphite are also denoted for comparison. It should be noted that all the data shown in this figure, are calculated by ourselves using the same method as T$_{5}$-carbon.}
\end{figure}

As shown in Fig. 1, T$_{5}$-carbon crystallizes in a fcc structure with a space group F$\overline{4}$3M (No. 216). The optimized lattice parameters of T$_{5}$-carbon are a = b =c = 5.574 $\textrm{\AA}$. One unit cell contains five carbon atoms, which occupy Wyckoff positions of 4a (0, 0, 0) and 16e (0.844, 0.844, 0.844). The denoted grey carbon ball in Fig. 1 connects three equivalent and an non-equivalent carbon neighbors, forming a quasi-$sp^{3}$ structure, and the blue ball is associated with four equivalent carbon atoms, constituting a standard $sp^{3}$ structure. There are two non-equivalent carbon-carbon bond lengths in T$_{5}$-carbon. The bond lengths are $d_{1}$= 1.490 $\textrm{\AA}$ and $d_{2}$= 1.501 $\textrm{\AA}$, respectively. There are three different bond angles among carbon atoms in T$_{5}$-carbon,  i.e., $\theta_1=60^{\circ}$, $\theta_2=109.471^{\circ}$ and $\theta_3=144.736^{\circ}$ as denoted in Fig. 1(a), respectively. Figure 2  plots the total energy per atom as a function of volume to confirm the energetic stability of T$_{5}$-carbon. Moreover, other carbon allotropes are also calculated for comparison using the same method. It can be seen that the stability of T$_{5}$-carbon in energy is slightly weaker than other carbon phases except for T-carbon, which is also confirmed by the calculated cohesive energy per atom, as shown in Table 1. These results reveal that T$_{5}$-carbon in energy is a comparable stable carbon allotrope against other carbon allotropes. To confirm the mechanical stability, we calculate the elastic constants of T$_{5}$-carbon. The calculated independent elastic constants $C_{11}$, $C_{12}$, and $C_{44}$ are 397.228, 158.929, and 195.368 GPa, respectively. These values meet the Born mechanical stability for cubic system\cite{BM-p}: $C_{11}$-$C_{12}$ $>$ 0 and $C_{11}$ + 2$C_{12}$ $>$ 0 and $C_{44}$ $>$ 0, confirming the mechanical stability of T$_{5}$-carbon. The phonon dispersion is also shown to confirm the dynamical stability of T$_{5}$-carbon. As shown in Fig. 3(a), there is no imaginary frequency existing in the phonon spectra, implying that T$_{5}$-carbon is dynamical stable. To further confirm the thermal stability of T$_{5}$-carbon, ab initio molecular dynamics simulation (AIMD) is implemented with a $3\times 3\times 3$ supercell. After a relaxation in 500 K for 10 ps with a time step of 1 fs, no structure change is observed. These results indicate that T$_{5}$-carbon is a comparable stable carbon phase, and thus may be synthesised in experiments.
\begin{table*}
\caption{\label{table:structure} The calculated equilibrium structural parameters including volume per atom $V_{0}$, lattice parameters $a$, $b$, and $c$, bond length $d$, cohesive energy per atom $E_{coh}$, bulk modulus $B_{0}$, and electronic band gap $E_{g}$ at zero pressure of Diamond, cR6, cT8, Graphite, Rh6, T-carbon and T$_{5}$-carbon, compared to experimental data \cite{experiment-GD-occelli2003}.}
\renewcommand\tabcolsep{3.0pt}
\begin{tabular}{c c c c c c c c }      %\tabnotes\noindent\begin{tabular}{@{}*{10}{1}}
\hline
\hline
   Structure                             &   Method                   &  $V_{0}(\textrm{\AA}^{3} /atom)$       &  a,b,c $(\textrm{\AA})$   &  d$(\textrm{\AA})$      &  $E_{coh} (\textrm{eV}/atom)$    & $B_{0}(GPa)$      &  $E_{g}(eV)$   \\
\hline
      Diamond ($F\overline{d}$3m)        & GGA                      &  5.69         & 3.572                     & 1.547               & 7.84    & 432      & 4.12   \\
                                         & Exp\cite{experiment-GD-occelli2003}     &  5.67         & 3.567                     & 1.544               &         & 446      & 5.47   \\
      cR6 ($R\overline{3}$m)             & GGA                      &  8.88         & 7.154,\ \ 3.602           & 1.354, 1.497        & 7.42    & 249      & 1.98   \\
      cT8 ($I4_{1}$/amd)                 & GGA                      &  8.49         & 5.931,\ \ 3.866           & 1.353, 1.495        & 7.46    & 275      & 2.54   \\
      Graphite ($p6_{3}$/mmc)            & GGA                      &  9.33         & 2.465,\ \ 6.886           & 1.424               & 8.04    & 276      &        \\
                                         & Exp\cite{experiment-GD-occelli2003}     &  8.78         & 2.460,\ \ 6.704           & 1.420               &         & 286      &        \\
      Rh6($R\overline{3} m$)             & GGA                      &  8.19         & 6.882,\ \ 3.595           & 1.357, 1.491        & 7.46    & 279      & 0.07   \\
      T-carbon ($F\overline{d}$3m)       & GGA                      &  13.25        & 7.520                     & 1.416, 1.500        & 6.67    & 159      & 2.24   \\
      T$_{5}$-carbon(F$\overline{4}$3M)  & This work                &  8.66         & 5.574                     & 1.501, 1.490        & 6.81    & 238      & 3.18   \\
\hline
\hline
\end{tabular}
\end{table*}

\begin{figure}\label{fig3}
\includegraphics[scale=0.46]{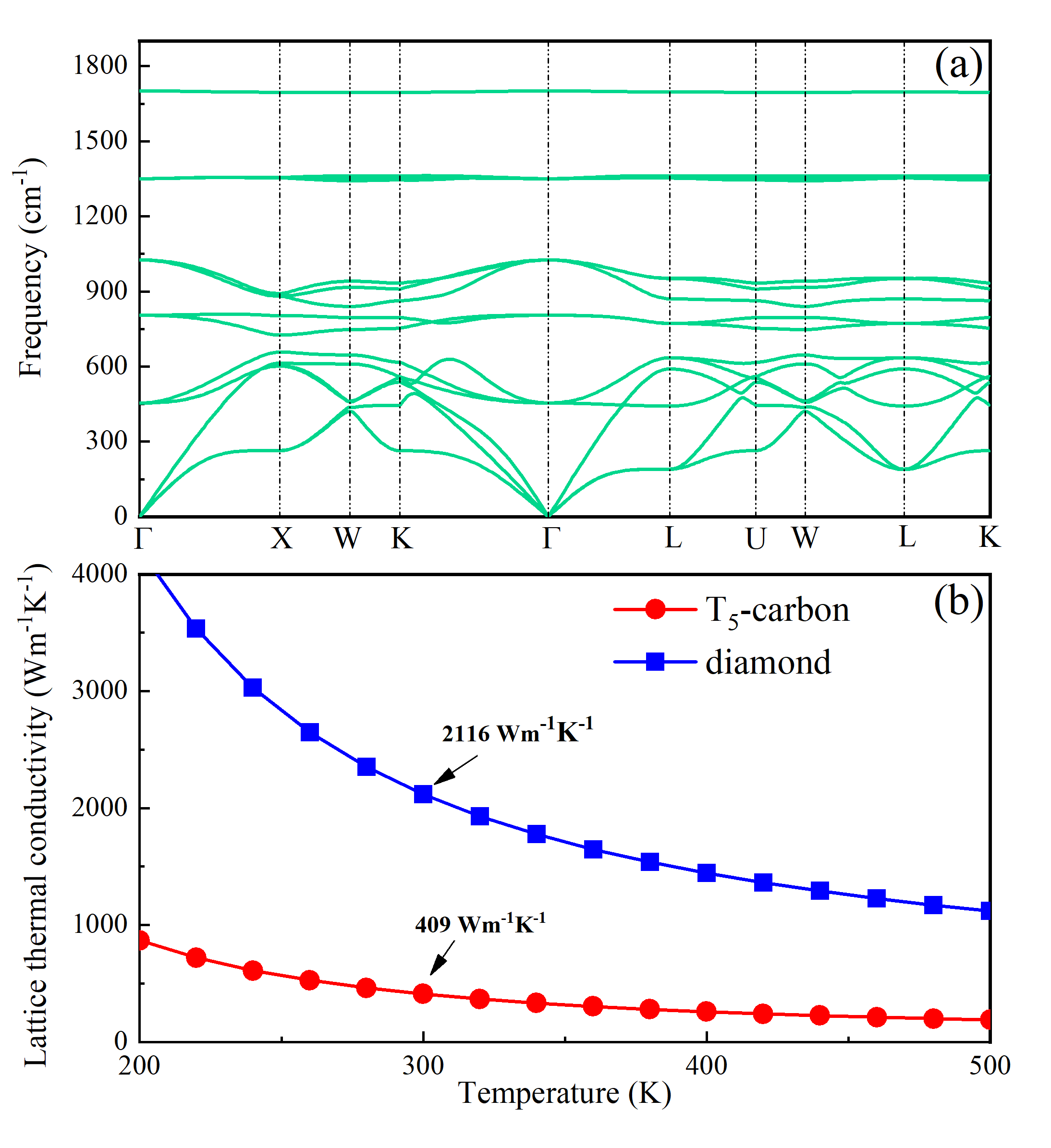}
\caption{(color online). (a) Phonon spectra of T$_{5}$-carbon. (b) Calculated lattice thermal conductivity versus temperature of diamond and T$_{5}$-carbon}
\end{figure}

\begin{figure}\label{fig4}
\includegraphics[scale=0.30]{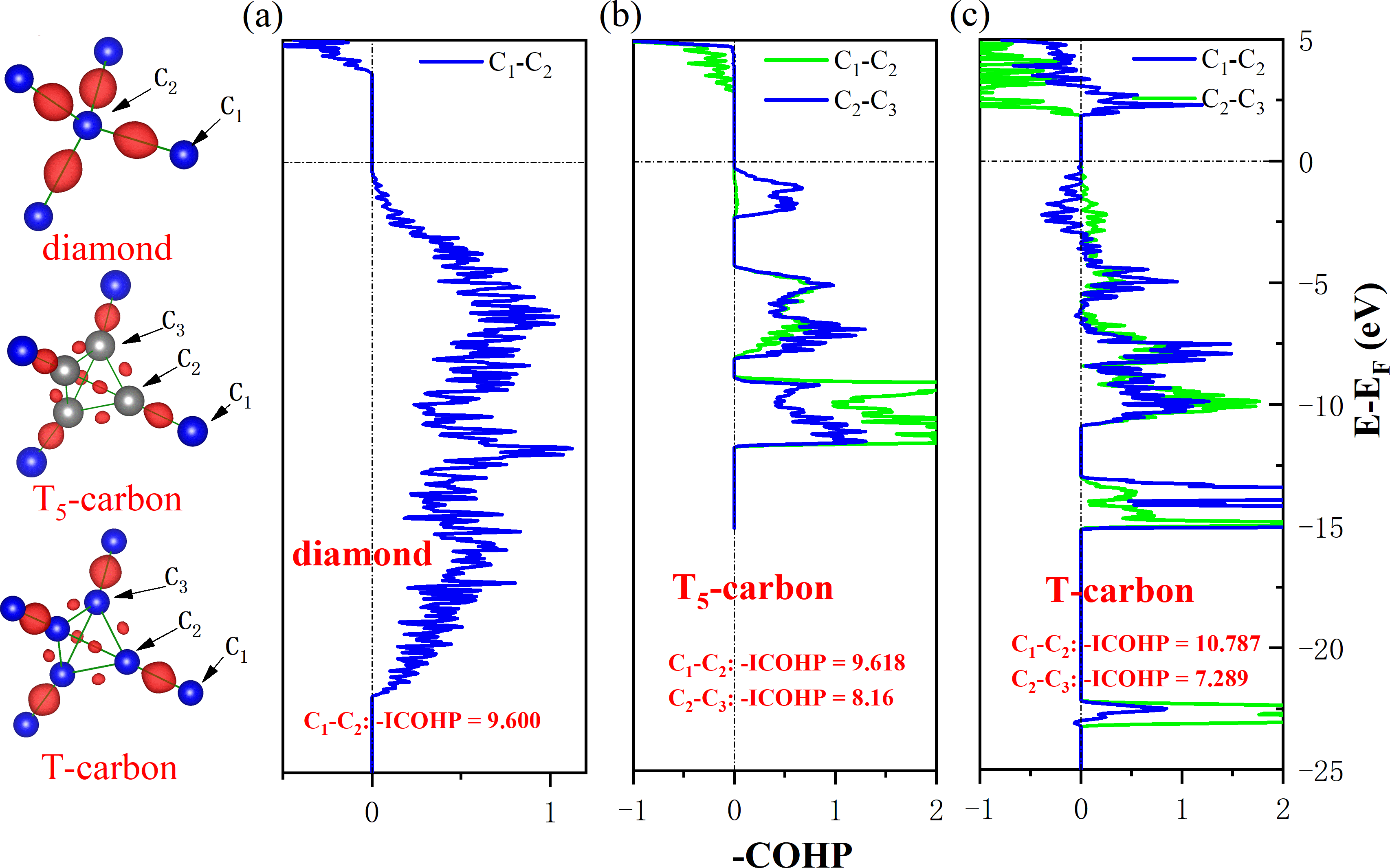}
\caption{(color online) Calculated COHP of diamond (a), T$_{5}$-carbon (b) and T-carbon (c), and the corresponding electronic charge density difference in the left panel.}
\end{figure}

Then, we investigate the mechanical and thermal properties of T$_{5}$-carbon. The following equation of Gao's model for vicker hardness calculation is used\cite{Gao2003}:
\begin{equation}\label{Hamiltonian}
H(GPa) = 350[(N_{e}^{2/3})e^{-1.191f_{i}}]/d^{2.5},
\end{equation}
where N$_{e}$ denotes the electron density of valence electrons per unit area, d denotes the bond length, and $f_{i}$ denotes the ionicity of the chemical bonds in a crystal measured by Phillips\cite{Phillips1970}. A Vicker hardness of 76.5 GPa is obtained by using this equation, which is bigger than T-carbon (61.1 Gpa) \cite{T-carbon} and c-BN (64.5 Gpa) \cite{Gao2003} and less than that of diamond (93.6 GPa)\cite{Gao2003}. To further investigate the mechanism of the comparable small Vicker hardness to diamond, the electron charge density difference and COHP of T$_{5}$-carbon are calculated and shown in Fig. 4. The corresponding results for diamond and T-carbon are also shown for comparison. The electron charge density differences suggest that the weaker Vicker hardness in T$_{5}$-carbon and T-carbon is owning to the weaker bonding strengths in the carbon tetrahedron, and the corresponding integral COHP (ICOHP) values also imply the same conclusion. We also investigate the thermal properties of the carbon tetrahedron by solving the phonon BTE, as shown in Fig. 3(b). The calculated lattice thermal conductivity for T$_{5}$-carbon is 409 W/mK at the temperature of 300 K. The obtained value is one order smaller than that of diamond (2116 W/mK)\cite{TC-carbonalltrope2017,TC-diamond1996} calculated by using the same method as T$_{5}$-carbon. Meanwhile, the value is one order bigger than T-carbon (33 W/mk)\cite{TC-carbonalltrope2017}, and the same order of magnitude as c-BN (768 W/mK) \cite{BN-TC}

\begin{figure}\label{fig5}
\includegraphics[scale=0.35]{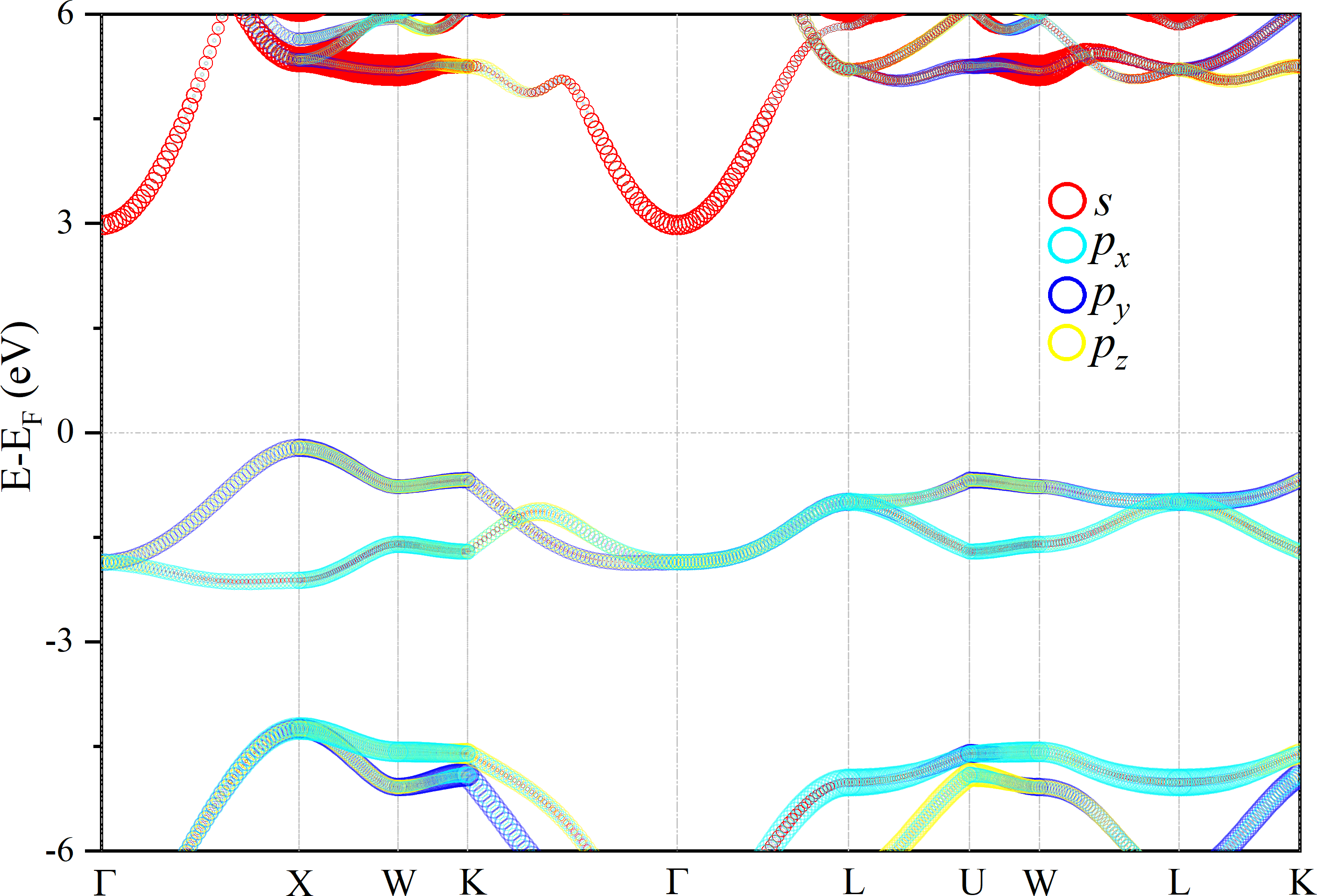}
\caption{(color online)  Projected band structure for T$_{5}$-carbon. The size of the red, cyan, blue and yellow circles represent the orbital $s$, $p_{x}$, $p_{y}$ and $p_{z}$.}
\end{figure}

Next, we calculate the electronic properties of T$_{5}$-carbon, and the results are shown in Fig. 5. The projected band structure reveal that the electronic states near the Fermi level are contributed by C-$s$ and C-$p$ orbital of carbon atoms, confirming the characters of hybridizations of C-$s$ and C-$p$. The electronic band structures reveal that T$_{5}$-carbon is a semiconductor with a indirect band gap of 3.18 eV. Furthermore, It is found that the band structures of valence band and conduction band have steep trend characters, which suggests a smaller effective mass. This result reveals that we may modify the band gap by doping for applying this carbon material to the field of photocatalysis.

\begin{figure}\label{fig6}
\includegraphics[scale=0.41]{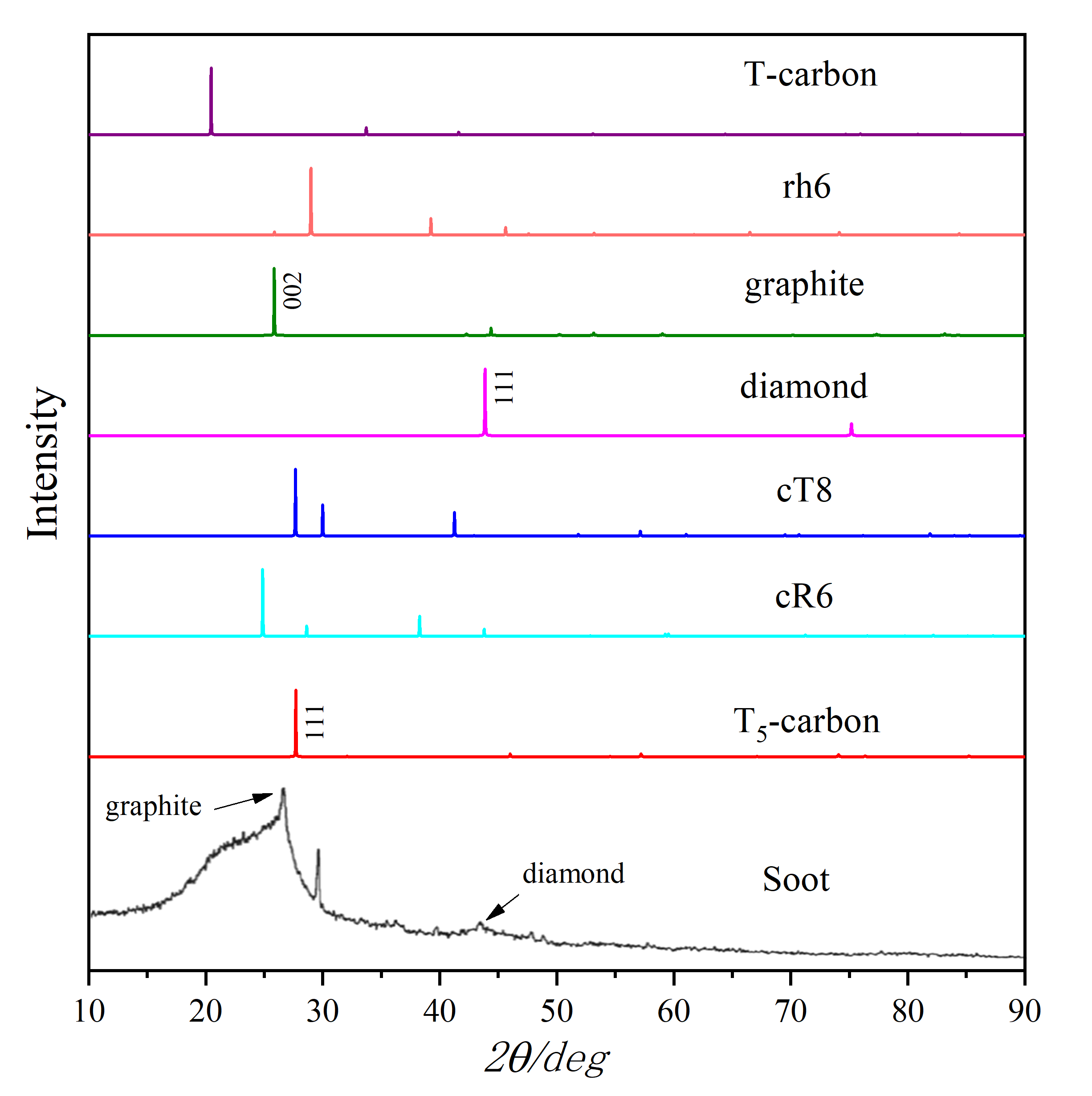}
\caption{(color online). Simulated XRD patterns for T-carbon, rh6, graphite, diamond, cT8, cR6 and T$_{5}$-carbon. Experimental XRD spectra of TNT and diesel oil detonation soot\cite{XRD-Soot2003}.}
\end{figure}

To provide more information for experimental observations, we simulate the XRD spctra of the carbon allotropes with an X-ray wavelength of 1.5406 $\textrm{\AA}$. As shown in Fig. 6, it can be seen that there is one sharp XRD peak at $2\theta$ = $27.69^{\circ}$ for (111) surfaces of the carbon tetrahedron. To further study the existence of the carbon tetrahedron by XRD patterns, we also compare its XRD spectra with previous experimental data from TNT and diesel oil detonation\cite{XRD-Soot2003}. The strong peak at $26.5^{\circ}$ is owning to graphite (002) diffraction and the weak peak around $43.5^{\circ}$ matches with the diamond (111) diffraction, confirming the existence of graphite and cubic diamond. Our simulated XRD spectrum on the surface of (111) of T$_{5}$-carbon is close to the strong peak of the Soot around $26.5^{\circ}$. The result may indicate the possible existence of T$_{5}$-carbon in this carbon Soot. In addition, the structure of T$_{5}$-carbon resembles that of T-carbon which is successfully synthesized by a pseudo-topotactic in methanol under picosecond pulsed-laser irradiation \cite{T-carbon-nonowire}. Therefore, we conclude that T$_{5}$-carbon may be experimentally synthesized by irradiating multi-walled carbon nanotubes with lasers.

In summary, a novel potential superhard carbon allotrope T$_{5}$-carbon is predicted by using first-principles calculations. The energetic, mechanical and thermal stabilities are checked by the calculations of energy-volume curve, elastic constants and AIMD simulations. The indirect band gap of T$_{5}$-carbon is 3.18 eV. The XRD spectrum is close to the experimental result of carbon Soot at 2$\theta$ = $26.5^{\circ}$. This novel carbon allotrope has a moderate lattice thermal conductivity of 409 W/mK. Importantly, the Vicker hardness of T$_{5}$-carbon is calculated to be 76.5 GPa, which is in the middle of diamond and c-BN.The result indicates T$_{5}$-carbon would have wide applications in superhard materials for mechanical devices.

~~~\\
~~~\\

This work was supported by the National Natural Science Foundation of China under Grant No. 11505003, the Science and Engineering Council through research grant (152-70-00017), the Open Project of State Key Laboratory of Environment-friendly Energy Materials under Grant No. 19kfhg03, the Chinese Academy of Sciences Hundred Talent Program under Grant No. Y2291810S3, the Provincial Natural Science Foundation of Anhui under Grant No. 1608085QA20, the Lift Engineering of Young Talents of Anhui University of Science and Technology, and use of computing resources at the A*STAR Computational Centre and National Supercomputer Centre, Singapore.\\
~~~\\

~~~\\
\textbf{Corresponding Authors}\\
$^\ast$Electronic address: chaozhang@mail.bnu.edu.cn\\

\end{document}